\documentclass[epj]{svjour}
\usepackage{graphics,epsfig}

\usepackage{xspace}

\newcommand{\lam}{\ensuremath{\Lambda} \xspace}

\newcommand{\kz}{\ensuremath{\mathrm{K^{0}_{S}}}\xspace}

\newcommand{\pt}{\ensuremath{p_{T}}\xspace}

\newcommand{\XI}{\ensuremath{\Xi^{-}}\xspace}

\newcommand{\pp}{\ensuremath{p+p}\xspace}
\newcommand{\ppbar}{\ensuremath{p+\bar{p}}\xspace}
\newcommand{\ee}{\ensuremath{e^{+}+e^{-}}\xspace}
\newcommand{\sqsRhic}{\ensuremath{\sqrt{s}=200}\xspace}
\newcommand{\sqsSps}{\ensuremath{\sqrt{s}=630}\xspace}

\title{``How important are next-to-leading order models in predicting
strange particle spectra in p+p collisions at STAR ?"}
\titlerunning{strange particles in p+p collisions vs pQCD models}

\authorrunning{M.Heinz}
\author{Mark Heinz\inst{1} for the STAR collaboration}
\offprints{mark.heinz@yale.edu}          
\institute{Physics Department, Yale University,  P.O. Box 208120,
New Haven, CT 06520-8120, USA}
\date{Received: date / Revised version: date}

\abstract{STAR has measured a variety of strange particle species in
\pp collisions at \sqsRhic GeV. These high statistics data are ideal
for comparing to existing leading- and next-to-leading order
perturbative QCD (pQCD) models. Next-to-leading (NLO) models have
been successful in describing inclusive hadron production using
parameterized fragmentation functions (FF) for quarks and gluons.
However, in order to describe identified strange particle spectra at
NLO, knowledge of flavor separated FF is essential. Such FF have
recently been parameterized using data by the OPAL experiment and
allow for the first time to perform NLO calculation for strange
baryons. In fact, comparing the STAR \lam data with these
calculations allow to put a constraint on the gluon fragmentation
function. We show that the Leading-order (LO) event generator PYTHIA
has to be tuned significantly to reproduce the STAR identified
strange particle data. In particular, it fails to describe the
observed enhancement of baryon-to-meson ratio at intermediate \pt
(2-6 GeV/c). In heavy-ion (HI) collisions this observable has been
extensively compared with models and shows a strong dependency on
collision centrality or parton density. In the HI context the
observed enhancement has been explained by recent approaches in
terms of parton coalescense and recombination models.}

\begin{document}

\maketitle

\setcounter{page}{1}

\section{Introduction}\label{intro}

Perturbative QCD has proven to be successful in describing inclusive
hadron production in elementary collisions. Within the theory's
range of applicability, calculations at next-to-leading order (NLO)
have produced accurate predictions for transverse momentum spectra
of inclusive hadrons at different energy scales
\cite{Borzumati:95,Marco:SQM04}. With the new high statistics
proton-proton data at \sqsRhic GeV collected by STAR, we can now
extend the study to identified strange hadrons as well as strange
resonances.

The perturbative QCD calculation applies the factorization ansatz to
calculate hadron production and relies on three ingredients. The
non-perturbative parton distribution functions (PDF) are obtained by
parameterizations of deep inelastic scattering data. They describe
quantitatively how the partons share momentum within a nucleus. The
second part, which is perturbatively calculable, consists of the
parton cross-section amplitude evaluated to LO or NLO using Feynman
diagrams. The third part consists of the non-perturbative
Fragmentation functions (FF) obtained from \ee collider data using
quark-tagging algorithms. These parameterized functions are
sufficiently well known for fragmenting light quarks, but less well
known for fragmenting gluons and heavy quarks. Recently, Kniehl,
Kramer and P\"otter (KKP) have shown that FF are universal between
\ee and \pp collisions \cite{KKP:01}.

The theoretical mechanisms of baryon production have been difficult
to understand and different attempts have been made
\cite{PythiaBaryon}. In the string fragmentation approach the
production of baryons is intimately related to di-quark production
from strings. They then combine with a quark to produce a baryon. In
NLO calculations, baryon production is based on the knowledge of
baryon fragmentation functions (FF) from \ee collisions. So far the
only baryon FF which has been accurately measured and parameterized
is that of the proton \cite{KKP:00}. Other groups have used a
statistical approach to calculate FF \cite{Bourrely}.

In the following section, we compare our p+p data to PYTHIA, the
most commonly used leading-order Monte Carlo event generator for
elementary collisions. In particular, we study predictions for
baryons and the ratios of baryons to mesons and see how parameter
tunes affect the data. We then compare our data with more
sophisticated NLO calculations.

\section{Data Analysis}\label{analysis}

The present data were reconstructed using the STAR detector system
which is described in more detail elsewhere \cite{STAR2}. The main
detector used in this analysis is the Time Projection Chamber (TPC)
covering the full acceptance in azimuth and a large pseudo-rapidity
coverage ($\mid \eta \mid < 1.8$). A total of 14 million non-singly
diffractive (NSD) events were triggered with the STAR beam-beam
counters (BBC) requiring two coincident charged tracks at forward
rapidity. Due to the particulary low track multiplicity environment
in p+p collisions, only 76\% of primary vertices are found
correctly; from the remainder, 14\% are lost and 10\% are badly
reconstructed as a MC-study showed. Of all triggered events, 7
million events passed the selection criteria requiring a valid
primary vertex within 50cm along the beam-line from the center of
the TPC. The strange particles were identified from their weak decay
to charged daughter particles. The following decay channels and the
corresponding anti-particles were analyzed: $\mathrm{K^{0}_{S}}
\rightarrow \pi^{+} + \pi^{-}$ (b.r. 68.6\%), $\Lambda \rightarrow p
+ \pi^{-}$(b.r. 63.9\%) ,$\Xi^{-} \rightarrow \Lambda +
\pi^{-}$(b.r. 99.9\%). Particle identification of the daughters was
achieved by requiring the dE/dx to fall within the 3$\sigma$-bands
of the theoretical Bethe-Bloch parameterizations. Further background
in the invariant mass was removed by applying topological cuts to
the decay geometry. Corrections for acceptance and particle
reconstruction efficiency were obtained, as a function of \pt, by a
Monte-Carlo based method of embedding simulated particle decays into
real events and comparing the number of simulated and reconstructed.

\begin{figure*}[t!]
\begin{center}
\includegraphics[width=15cm]{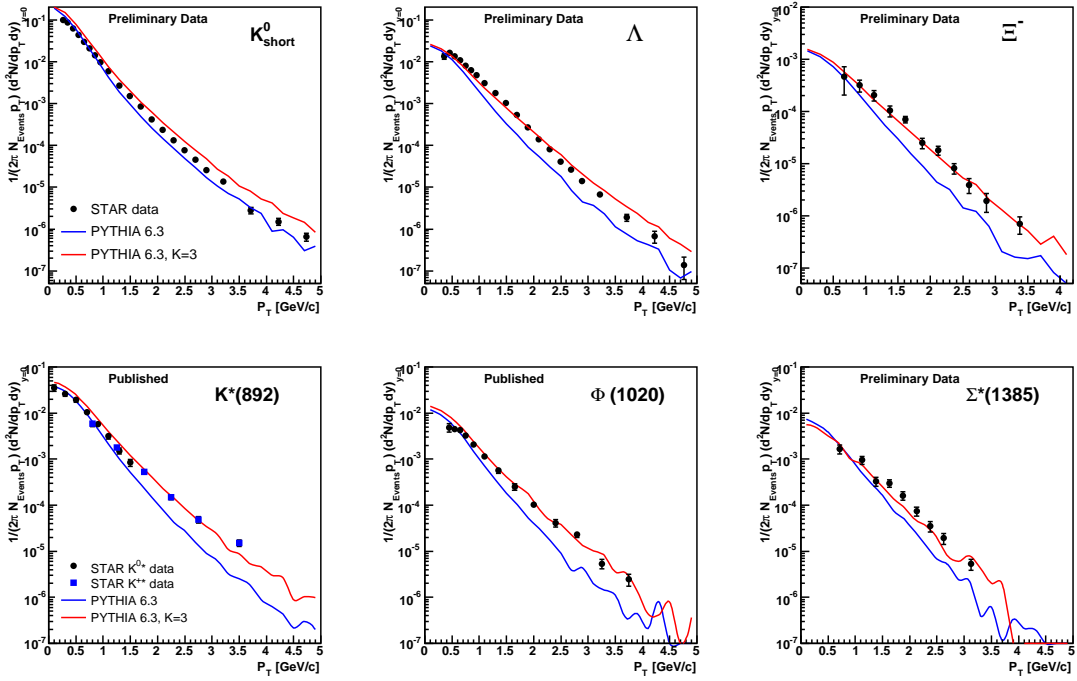}
\caption{(Top) Minimum-bias \pt~spectra for \kz, \lam and $\Xi$ at
($\mid{y}\mid < 0.5$) from $p+p$ at $\sqrt{s} = 200 $GeV. (Bottom)
$K^{*}$ and $\phi$, and $\Sigma^{*}$ \pt~spectra at mid-rapidity. In
the left panel, black symbols are $K^{0*}$ and blue symbols are
$K^{+*}$ \cite{Kstar:05,Phi:05,Markert:06}. } \label{fig:strangeLO}
\end{center}
\end{figure*}

\section{Comparison to PYTHIA}\label{pythia}
\subsection{Strange particle spectra}

One of the most widely used models for simulating elementary
collisions is PYTHIA \cite{Pythia:87}. It is a parton-shower based
event generator that includes leading order parton processes and
parton fragmentation based on the Lund Model. The parton
distributions of the initial state protons can be chosen from an
array of PDFs (here we use CTEQ5M). The model is being actively used
and the authors have recently released a version with completely
overhauled multiple scattering and shower algorithms (version 6.3)
\cite{Pythia:04}. The PYTHIA version used in this paper is 6.317.

The string fragmentation based on the Lund Model requires only two
parameters to define the shape of the fragmentation function and is
universal for all light quark flavors. Baryons are produced from
di-quarks and their probability is suppressed with respect to
$\bar{q}q$ pair production. Next-to-leading order processes can be
``simulated" in PYTHIA by tuning the K-factor (MSTP(33)) or by
increasing the parton shower activity. This will enhance the
relative probability of hard processes of type quark-gluon and thus
mock-up the contributions from higher order processes.

In figure \ref{fig:strangeLO} (upper row), we compare PYTHIA
calculations for strange mesons and baryons to the measured STAR
data. Whereas the default parameters (blue line) agree quite well
for the \kz, they clearly underestimate the yields at intermediate
\pt for the \lam and \XI. By increasing the K-factor to 3 (red line)
we achieve a reasonable agreement with the data. In figure
\ref{fig:strangeLO} (lower row), we compare PYTHIA to the strange
resonances $K^{*}$, $\phi$ and $\Sigma^{*}$. Again, only when
applying a higher K-factor does the calculation agree with our data.

In summary, PYTHIA is capable of describing \pt spectra for a
variety of particles from $p+p$ collisions at RHIC energies.
However, we have presented evidence that a tune of the LO K-factor
is necessary in particular for strange baryons and resonances. Of
course, we have not explored all possibilities of parameter ``tunes"
and there may be other, equivalent ways of reproducing the data.

What are the possible reasons for this discrepancy? The ``naive"
reason, supported by the K-factor tune, is that higher order
contributions may be significant. However it is troubling that the
pions do not seem to require this tune as shown previously
\cite{Heinz:wwnd}, even though a similar study of K-factors for
non-identified hadrons found that at $\sqrt{s} = 200$GeV a value of
3 was needed \cite{Eskola:03}.

Another, perhaps more natural explanation, may be related to
fragmentation functions for baryons in PYTHIA. In the next section
we will discuss possible changes to the baryon production
parameters, ie. the di-quark suppression factors, which may help
solve this discrepancy.

\subsection{Baryon production}
\begin{table}
\caption{STAR dN/dy for various baryons from \pp collisions at
\sqsRhic~GeV ($|y|<$0.5) compared to PYTHIA 6.317. Pythia Baryon
tune is defined as PARJ(1)=0.125 (D=0.1) and PARJ(3)=0.5 (D=0.4).}
\label{tab:baryonTable}       
\begin{tabular}{llll}
\hline\noalign{\smallskip}
Particle & STAR dN/dy & PYTHIA & PYTHIA tuned  \\
\noalign{\smallskip}\hline\noalign{\smallskip}
proton   & 0.11 $\pm$ 0.01     & 0.096  & 0.11   \\
\lam (FD)& 0.0385 $\pm$ 0.0035 & 0.0297 & 0.0371 \\
\XI      & 0.0026 $\pm$ 0.0009 & 0.0020 & 0.0029 \\
\noalign{\smallskip}\hline
\end{tabular}
\end{table}

In string models, baryon production in its simplest form is
understood via the production of di-quark pairs from string-breaking
and their recombination with other quarks. This process is
suppressed with respect to $\bar{q}-q$ pairs from string-breaking
resulting in systematically lower baryon yields than mesons. The
default value for the suppression factor is $P(qq)=0.1\times P(q)$.
We have increased this value to 0.125 (PYTHIA parameter PARJ(1)).
Similarly the strange di-quarks are suppressed with a default factor
$P(sq)=0.4\times P(q)$, which we have increased to 0.5 (PYTHIA
parameter PARJ(3)).

In table \ref{tab:baryonTable} we show recently measured baryon
yields in \pp collisions at mid-rapidity. Values for \lam have been
corrected for feed-down (FD) from \XI-decays. From the values in the
table, it is clear that the tuned values for PYTHIA are in better
agreement with the experimental measurements of STAR than the
default values. However, it must be said that the agreement is
confined to low \pt and that this tune does not change the shape of
the PYTHIA spectra to improve the high \pt part.

\subsection{Baryon to Meson ratios}

Recent heavy-ion data from STAR show a large enhancement of the
baryon to meson ratios at intermediate \pt, which is associated with
parton coalescence and recombination models \cite{Lamont:06}. \lam
and \kz are ideal candidates for comparing baryon to meson
production at these momenta since they can be cleanly identified via
the topological reconstruction method described at the beginning.

\begin{figure}[t!]
\includegraphics[width=8cm]{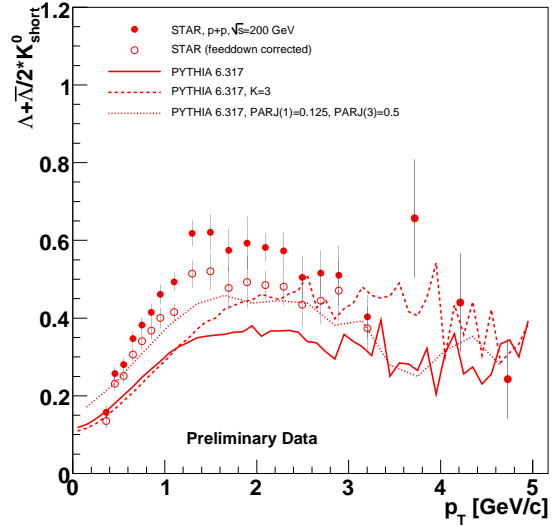}
\caption{Ratio of $\lam+\bar{\lam}/2\times \kz$ vs \pt from STAR
data compared to three different tunes of PYTHIA 6.317. Data are
shown with and without feed-down correction, whereas PYTHIA
calculations are corrected for feed-down.} \label{fig:bmratio1}
\end{figure}

In Figure \ref{fig:bmratio1} we show the measured $\lam/\kz$ ratio
vs \pt measured by STAR, together with 3 different calculations by
PYTHIA. Open symbols depict \XI feed-down corrected \lam yields.
Clearly, the default PYTHIA calculation lies well below the data.
Increasing the LO K-factor does not improve the ratio much at low
\pt, although it does describe the ratio at high \pt. However, using
the tuned baryon parameters discussed in the previous section
improves the agreement at low \pt considerably. Thus, we need to us
a combination of both K-Factor and baryon parameter tune to
simultaneously describe the spectra and the ratios.

\begin{figure}[t!]
\includegraphics[width=8cm]{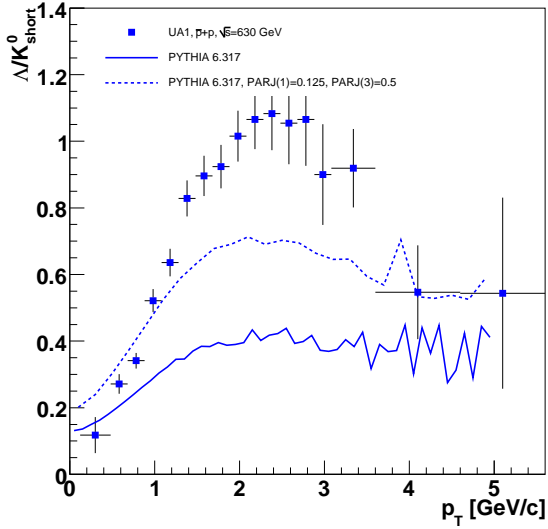}
\caption{Ratio of $\lam/\kz$ vs \pt from UA1 data compared to two
different tunes of PYTHIA 6.317 } \label{fig:bmratio2}
\end{figure}

This result triggers the interesting question as to the possible
energy dependence of this baryon production parameter. To
investigate this further, we have used data for strange mesons and
baryons from the UA1 collaboration, which measured \ppbar collisions
at \sqsSps~GeV and produced the particle ratio presented in figure
\ref{fig:bmratio2} \cite{UA1}.

The figure clearly shows that the disagreement with default PYTHIA
for the baryon to meson ratio is not specific to our energy scale
but also exists at higher energies. At \sqsSps~GeV, the difference
between PYTHIA and data is about a factor of 3 and the enhancement
of $\lam/\kz$ is twice as large as in STAR. Even when tuning PYTHIA
to the same values as for STAR the discrepancy between data and
model remains large. This may be an indication that the effects
observed in this ratio in heavy-ion data are present in some form in
\pp data. It remains to be understood whether the enhancement of the
ratio is due to parton density (multiplicity) or to collision
energy.

\section{Comparison to next-to-leading order pQCD}\label{nlo}

In this final section we discuss the improvements which have
recently been made by next-to-leading order calculations using more
precisely parameterized fragmentation functions. Fragmentation
functions for separated quark flavors have been notoriously
difficult to obtain due to the lack of sufficiently precise collider
data. However, OPAL has recently published flavor tagged data from
\ee collisions which allowed theorists to compute better
fragmentation functions \cite{OPAL:00}.

\begin{figure}[h]
\includegraphics[width=8cm]{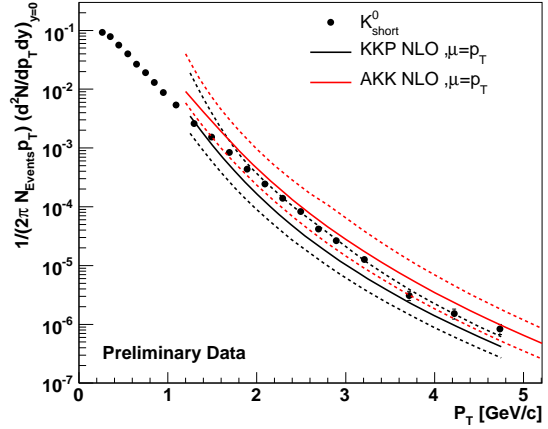}
\caption{\pt~spectra for \kz at midrapity ($\mid{y}\mid < 0.5$) from
$p+p$ at $\sqrt{s} = 200 $GeV compared to two different NLO
calculations. Dashed lines indicate the scale uncertainty of the NLO
calculation, ie. $\mu=0.5\pt$ (lower), $\mu=2\pt$ (upper). }
\label{fig:k0NLO}
\end{figure}

\begin{figure}[h]
\includegraphics[width=8cm]{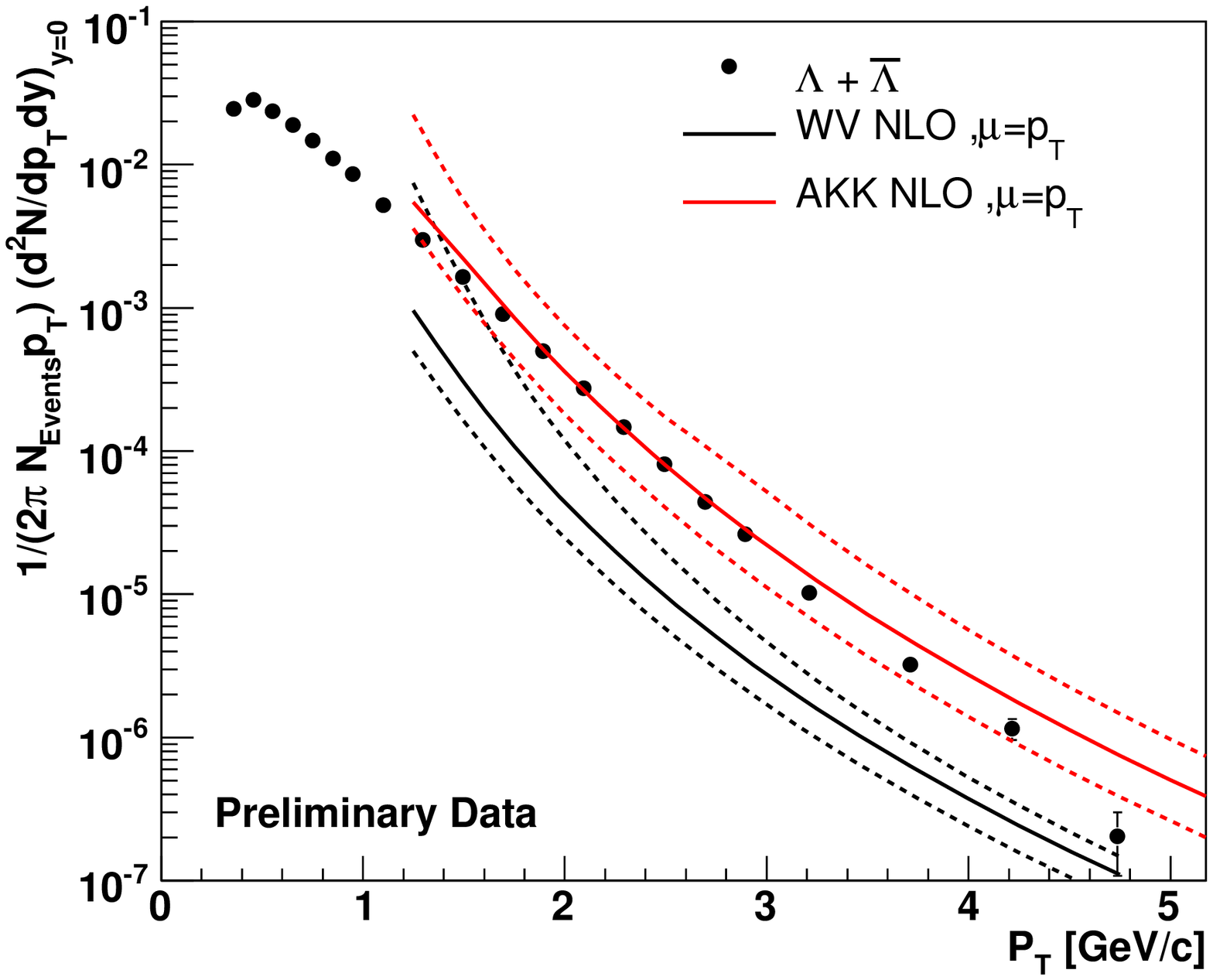}
\caption{\pt~spectra for \lam at midrapity ($\mid{y}\mid < 0.5$)
from $p+p$ at $\sqrt{s} = 200 $GeV compared to two different NLO
calculations.} \label{fig:lamNLO}
\end{figure}

In figures \ref{fig:k0NLO} and \ref{fig:lamNLO} we compare two
different NLO calculations to our \kz and \lam data . The first one
(black lines) uses older FF by Kniehl \emph{et al.} (KKP) and
Vogelsang \emph{et al} (WV) \cite{deFlorian:PRD57}. The second one
(red lines) was done by Albino \emph{et al.} (AKK) using more recent
FF based on the light flavor tagged OPAL data \cite{AKK:06}.

Clearly, these newer parameterizations improve the description of
our \lam~data greatly. However, in order to achieve this agreement,
they fix the initial gluon to \lam fragmentation function
($D_{g}^{\Lambda}$) to that of the proton, then estimate that an
additional scaling factor of 3 is necessary to achieve agreement
with STAR data. However, this modified FF for $D_{g}^{\Lambda}$ also
works well in describing the $p+\bar{p}$ SPS data at $\sqrt{s} =
630$GeV. It therefore appears that the STAR data is a better
constraint for the high z part of the gluon fragmentation function
than the OPAL \ee data. Similar conclusions with respect to the
important role of \pp collisions have been drawn elsewhere
\cite{Levai:PRL02}.

\section{Summary}

We have shown that the theoretical description of identified strange
particles in \pp and \ppbar collisions is still not fully
understood. This is especially important since these models are now
extensively used to predict observables for the LHC-era, and
therefore one should be aware of their limitations. Phenomenological
LO models can be tuned to describe the data but still struggle to
describe baryon production at intermediate \pt.

Baryon production, and in particular the baryon to meson yield
ratios at intermediate \pt, are one of the ``hot" topics in current
heavy-ion research at RHIC. The \pp data presented here allows us to
look at the ratio in elementary collisions and check how well it is
understood in a simple system. The fact that PYTHIA baryon
production parameters need to be tuned quite considerably to achieve
an agreement is interesting. We also showed that at the higher
energies, i.e. \sqsSps GeV, this difference is even larger. This is
an indication that the baryon to meson effects previously observed
in heavy-ion collisions are present in some form in \pp data, and
that the associated physics phenomena therefore need to be explained
without requiring the presence of a quark-gluon plasma.

Next-to-Leading order calculations have greatly improved with light
flavor tagged fragmentation functions. However the high-z range of
the gluon FF previously extracted from \ee data seems inconsistent
with \pp and \ppbar data, indicating that RHIC data could be
valuable in constraining the gluon FF.

\section*{Acknowledgments}
The author would like to acknowledge theoretical calculations and
enlightening discussions with Simon Albino (AKK) and Peter Skands
(PYTHIA).

We thank the RHIC Operations Group and RCF at BNL, and the NERSC
Center at LBNL for their support. This work was supported in part by
the Offices of NP and HEP within the U.S. DOE Office of Science; the
U.S. NSF; the BMBF of Germany; CNRS/IN2P3, RA, RPL, and EMN of
France; EPSRC of the United Kingdom; FAPESP of Brazil; the Russian
Ministry of Science and Technology; the Ministry of Education and
the NNSFC of China; IRP and GA of the Czech Republic, FOM of the
Netherlands, DAE, DST, and CSIR of the Government of India; Swiss
NSF; the Polish State Committee for Scientific Research; SRDA of
Slovakia, and the Korea Sci. and Eng. Foundation.

\vfill\eject
\end{document}